\documentclass[aps,pre,groupaddress,superscriptaddress,reprint,amsmath,amssymb,graphicx]{revtex4-1}
\usepackage{graphicx}
\usepackage{physics}
\usepackage{bm}
\usepackage{bbold}
\usepackage{xcolor}

\begin{document}
\title{Mutual information in changing environments: non-linear interactions, out-of-equilibrium systems, and continuously-varying diffusivities}
\author{Giorgio Nicoletti}
\affiliation{Laboratory of Interdisciplinary Physics, Department of Physics and Astronomy ``G. Galilei'', University of Padova, Padova, Italy}
\author{Daniel Maria Busiello}
\affiliation{Institute of Physics, \'Ecole Polytechnique F\'ed\'erale de Lausanne - EPFL, 1015 Lausanne, Switzerland}

\begin{abstract}
    Biochemistry, ecology, and neuroscience are examples of prominent fields aiming at describing interacting systems that exhibit non-trivial couplings to complex, ever-changing environments. We have recently shown that linear interactions and a switching environment are encoded separately in the mutual information of the overall system. Here, we first generalize these findings to a broad class of non-linear interacting models. We find that a new term in the mutual information appears, quantifying the interplay between non-linear interactions and environmental changes, and leading to either constructive or destructive information interference. Furthermore, we show that a higher mutual information emerges in out-of-equilibrium environments with respect to an equilibrium scenario. Finally, we generalize our framework to the case of continuously varying environments. We find that environmental changes can be mapped exactly into an effective spatially-varying diffusion coefficient, shedding light on modeling and information structure of biophysical systems in inhomogeneous media.
\end{abstract}

\maketitle

\section{Introduction}
An accurate description of real-world systems should capture both their internal interactions and their couplings with noisy, ever-changing environments. The main difficulty stems from the fact that often environmental changes are not directly observable, hence leading to the necessity of more simplified yet informative approaches. The more simplistic one might be to ignore environmental effects. However, it is now well-understood that these are fundamental ingredients in many different fields, from biology to neuroscience \cite{Swain2002, Zhu2009, hilfinger2011separating, Bowsher2012, Tsimring2014, Thomas2014, dass2021equilibrium, Mariani2021}. A slightly more complete understanding of real-world systems would come from the estimation of effective couplings, in principle affected also by the presence of a changing environment \cite{schneidman2006weak,mora2010maximum,bialek2012statistical}. Although this idea might lead to descriptive models, it makes impossible to understand whether observed behaviors originate from the internal interactions or are sheer consequences of a shared environment. 

In this intricate scenario, information theory might be the leading framework to determine the role of different coupling sources in shaping complex systems' behaviors. In particular, a key quantity is the mutual information associated with two stationary processes $x_1(t)$ and $x_2(t)$, 
\begin{align}
\label{eqn:mutual}
    I & = \int dx_1dx_2 \, p(x_1, x_2) \log \frac{p(x_1, x_2)}{p(x_1) p(x_2)}
\end{align}
which is nothing but the Kullback-Leibler divergence between $p(x_1, x_2)$, the joint stationary probability distribution, and $p(x_1)p(x_2)$, the product of their marginalized stationary distributions \cite{ThomasCover2006}. $I$ quantifies the overall dependency between $x_1$ and $x_2$. Recently, we showed that the mutual information of systems with linear interactions and a switching discrete-state environment receive disentangled contributions from environmental and internal interactions \cite{nicoletti2021mutual}. This result revealed that the properties of the information content of complex systems can be particularly informative, and that tools from information theory may greatly help to capture their essential features \cite{nicoletti2022information}.

Here, we generalize our previous results to more complex systems, highlighting criticalities and potentialities of the proposed approach. First, we show that the presence of non-linear interactions may give rise to a new interference term in the mutual information. This additional contribution may lead to both an increase and a decrease of the mutual information with respect to the sum of the contributions associated with the environment and the internal interactions. A careful analysis of the system under investigation here reveals the phenomenological origin of this constructive or destructive information interference. 

Then, we show that in systems placed out-of-equilibrium by the presence of a multiplicative noise (e.g., a thermal gradient), in the absence of interactions, the environmental information increases with the magnitude of the non-equilibrium term. Finally, we consider the case in which the environment is described by a continuous process \cite{Chechkin2017,Wang2020}. We show that the effect of the environment at stationarity can be mapped into a heterogeneous diffusion coefficient, i.e., an effective inhomogeneous media. Thus, the presence of a changing environments cannot always be mapped into effective interactions among degrees of freedom, as one may naively believe, but sometimes it manifests into indirect spatial couplings. This result highlights a potential warning for inference methods trying to estimate interactions from measured data.

\section{Time-scale separation approach}
Consider an interacting system of $N$ possibly interacting particles that share the same changing environment, whose effect is to modify the overall diffusion coefficient. In general, we assume that we have a finite number $M$ of environmental states, i.e., the diffusion coefficient of the system only takes discrete values. We will eventually relax this condition.

This framework is analogous to the one introduced in \cite{nicoletti2021mutual}. It is described by the following Fokker-Planck equation,
\begin{align}
    \label{eqn:fokker_planck}
    \partial_t p_i(\vb{x}, t) = & \sum_{\mu = 1}^N\partial_\mu\left[F_\mu(\vb{x})p_i(\vb{x}, t)\right] + \sum_{\mu= 1}^N\,\partial_\mu^2 \left[D_ip_i(\vb{x}, t)\right] +\nonumber\\
    & + \sum_{j = 1}^M \left[W_{j \to i} p_j(\vb{x}, t) - W_{i \to j} p_i(\vb{x}, t)\right],
\end{align}
where $\vb{x} = (x_1, \dots, x_N)$ indicates all internal degrees of freedom, $W_{i \to j}$ is the transition rate from the $i$-th to the $j$-th environmental state, and $D_i$ is the diffusion coefficient associated with such states. We are interested in the stationary solution of Eq.~\eqref{eqn:fokker_planck}, whose finding is, in general, a particularly challenging task. Therefore, we resort to a time-scale separation approach, in which the environment can be either much faster or much slower than all timescales at which the internal dynamics operates.

We name $\tau$ the fastest internal timescale, whereas the jump process between the environmental states happens at a typical timescale $\tau_{\rm env}$. Let us first consider the limit $\tau/\tau_{\rm env} := \delta \ll 1$ and seek a formal solution of the form
\begin{equation}
    \label{eqn:expansion_slow_jumps}
    p_i(\vb{x}, t) = p_i^{(0)}(\vb{x}, t) + \delta \, p_i^{(1)}(\vb{x}, t) + \mathcal{O}(\delta^2).
\end{equation}
Re-scaling the time by the slowest timescale, that is $t \to t/\tau_{\rm env}$, we end up with
\begin{align}
    \partial_t p_i^{(0)} = & \frac{1}{\delta}\sum_{\mu = 1}^N \biggl[\partial_\mu\left[\tilde{F}_\mu(\vb x)p_i^{(0)}\right] + \partial_\mu^2 \left[ \tilde{D}_i p_i^{(0)} \right]\biggl]+ \nonumber\\ 
    & + \sum_{\mu = 1}^N \biggl[\partial_\mu\left[\tilde{F}_\mu(\vb x)p_i^{(1)}\right] + \partial_\mu^2 \left[\tilde{D}_i p_i^{(1)} \right]\biggl]+ \nonumber \\ 
    & + \sum_{j = 1}^M \left[\tilde W_{j \to i} p_j^{(0)} - \tilde W_{i \to j} p_i^{(0)}\right] + \mathcal{O}(\delta) \nonumber
\end{align}
where $\tilde F_\mu := \tau F_\mu$, $\tilde D_i := \tau D_i$, and $\tilde W_{i \to j} := \tau_{\rm env} W_{i \to j}$. The leading $\delta^{-1}$ order corresponds to the stationary solution of the Fokker-Planck equation associated with the fastest dynamics alone. Here, this is equal to the distribution $P_i^{\rm st}(\vb{x})$ that solves the interacting dynamics at a fixed environmental state $D_i$:
\begin{equation}
\label{eqn:mixture_components}
    0 = \sum_{\mu = 1}^N \biggl[\partial_\mu\left[F_\mu(\vb x)P_i^{\rm st}(\vb{x})\right] + \partial_\mu^2 \left[D_i P_i^{\rm st}(\vb{x}) \right] \biggl].
\end{equation}
We can always assume that the zero-th order solution of Eq.~\eqref{eqn:fokker_planck} can be written as $p_i^{(0)}(\vb{x}, t) = \pi_i(t)P_i^{\rm st}(\vb{x})$ (see also \cite{busiello2020coarse}). Then, by integrating over $\vb{x}$, the $\mathcal{O}(1)$ order gives
\begin{align*}
    \partial_t \pi_i(t) = \sum_{j = 1}^M \left[\tilde W_{j \to i} \pi_j(t) - \tilde W_{i \to j} \pi_i(t)\right] .
\end{align*}
Hence, the zero-th order for the steady state reads as follows
\begin{align}
    \label{eqn:solution_slow_jumps}
    p^{(0)}_i(\vb{x})|_{\delta \ll 1} = p_{\rm slow}(\vb{x}) = \sum_{i = 1}^M \left[\pi_i P_i^{\rm st}(\vb{x})\right]
\end{align}
where $\pi_i$ are the stationary probabilities of the jumps process alone, and the subscript “slow" refers to the fact the environment is the slowest process in this limit. Eq.~\eqref{eqn:solution_slow_jumps} is a mixture distribution, where the mixture components are the stationary solutions obtained with a fixed environmental state $i$, $P_i^{\rm st}(\vb{x})$.

These calculations can be easily carried out in the opposite limit, $\tau_{\rm env}/\tau = \delta^{-1} \ll 1$, i.e., when the environment is much faster than the internal processes. In this case, the stationary joint probability distribution $p_{\rm fast}(\vb{x})$ that solves Eq.~\eqref{eqn:fokker_planck} is given by the solution of
\begin{align}
    \label{eqn:solution_fast_jumps}
    0 = & \sum_{\mu = 1}^N \left[\partial_\mu\left[F_\mu(\vb x) p_{\rm fast}(\vb{x})\right] + \partial_\mu^2 \left[\left(\sum_i\pi_i D_i\right)p_{\rm fast}(\vb{x}) \right]\right].
\end{align}
The system feels an effective diffusion coefficient which is the ensemble average of all environmental states, as a consequence of the presence of the environment.

\section{Mutual information: dependencies and bounds}

For the sake of simplicity, here we focus on the case of two particles moving in a $1D$ space, $\vb{x} = (x_1,x_2)$, and two environmental states specified by the diffusion coefficients $D_-$ and $D_+$. The multidimensional generalization is straightforward. In \cite{nicoletti2021mutual}, we showed that, in the presence of linear interactions - i.e., when Eq.~\eqref{eqn:solution_slow_jumps} corresponds to a Gaussian mixture - the mutual information can be exactly disentangled into two independent contributions. The first one depends solely on the environmental dimensionless parameters, $D_-/D_+$ and $w_-/w_+$, where $w_\pm$ is the rate of transition into the state $i = \pm$. The second, instead, only depends on the internal interactions between the two particles. Hence, the dependencies between $x_1$ and $x_2$ induced by the environment and by the internal interactions are fully disentangled. 

We now relax the assumption of linear interactions and explore the effects of non-linear couplings. In particular, we already know that the mutual information in the fast-jumps limit, Eq.~\eqref{eqn:solution_fast_jumps}, only contains the contribution from the internal interactions, since the environment results into a constant effective diffusion. Conversely, the slow-jumps limit is much more intriguing. In this case, both environment and internal couplings will contribute to the mutual information between $x_1$ and $x_2$, $I(x_1,x_2)$, but their interplay is far from being trivial and expected.

As a general remark, we notice that the mutual information can only depend on dimensionless quantities. These, in turn, may depend on environmental features, internal parameters, or combinations of both. In the slow-jumps limit, by inspecting Eq.~\eqref{eqn:solution_slow_jumps}, the parameter $w_+/w_-$ can only enter through $\pi_\pm = w_\pm/(w_\pm + w_\mp)$. Then, the stationary solution of the dynamics at a fixed environment, $P_\pm^{\rm st}$, determines all the other dimensionless parameters in play.

Moreover, the slow-jumps limit allows us to consider some simple bounds \cite{ThomasCover2006} on the entropy of a mixture distribution, and thus on the mutual information. Let $H_{12}^i$ be the entropy of the $i$-th component $P_i^{\rm st}(x_1, x_2)$ of the joint distribution. This entropy is bounded by
\begin{equation}
    \sum_i \pi_i H_{12}^i \le H_{12} \le \sum_i \pi_i \left[H_{12}^i - \log \pi_i\right].
\end{equation}
An analogous bound can be cast for $H_{1}^i$ and $H_{2}^i$, respectively the entropies of the $i$-th component $P_i^{\rm st}(x_{1})$ and $P_i^{\rm st}(x_{2})$ of the marginal distributions. Therefore, the mutual information of the mixture distribution, Eq.~\eqref{eqn:solution_slow_jumps}, is bounded by
\begin{equation}
\label{eqn:bound_mutual}
    \sum_i \pi_i I^i_{12} - H_{\rm jumps} \le I_{12} \le \sum_i \pi_i I^i_{12} + 2 H_{\rm jumps}
\end{equation}
where $H_{\rm jumps} = -\sum_i\pi_i \log\pi_i$ is the entropy associated with the environmental jumps, and $I^i_{12} = H_1^i + H_2^i - H^i_{12}$ is the mutual information associated with the $i$-th component $P_i^{\rm st}(x_1, x_2)$ of the joint distribution.

These bounds can be greatly improved \cite{Kolchinsky2017, nicoletti2021mutual}, provided our ability to compute some suitable information distances both between the components of the mixture distribution in Eq.~\eqref{eqn:solution_slow_jumps} and the components of the two corresponding marginalization. However, besides the Gaussian case, this is often challenging. Generally speaking, Eq.~\eqref{eqn:bound_mutual} shows that the mutual information cannot be larger than the sum of the weighted average of the mutual information in the different environmental states and twice the entropy of the jumps. Albeit loose, this upper bound shows that, in principle, the system may contain more information than the sum of the contributions stemming from the environment and the internal interactions. Therefore, on the one hand, we expect and later show that the presence of non-linear interactions might undermine the exact disentangling holding for the linear case. On the other hand, in what follows we also report situations in which the presence of non-linearities boost, or even suppress, the overall mutual information due to internal and environmental couplings. This observations effectively hinder our ability to pinpoint the presence of interactions in complex systems, but reveal surprising properties about their information structure.

%slow-jumps and fast-jumps limit, adimensional dependencies of the mutual information, simple bounds.

\section{Environmental contribution with non-linear relaxation}
\begin{figure*}[t]
    \centering
    \includegraphics[width=0.98\textwidth]{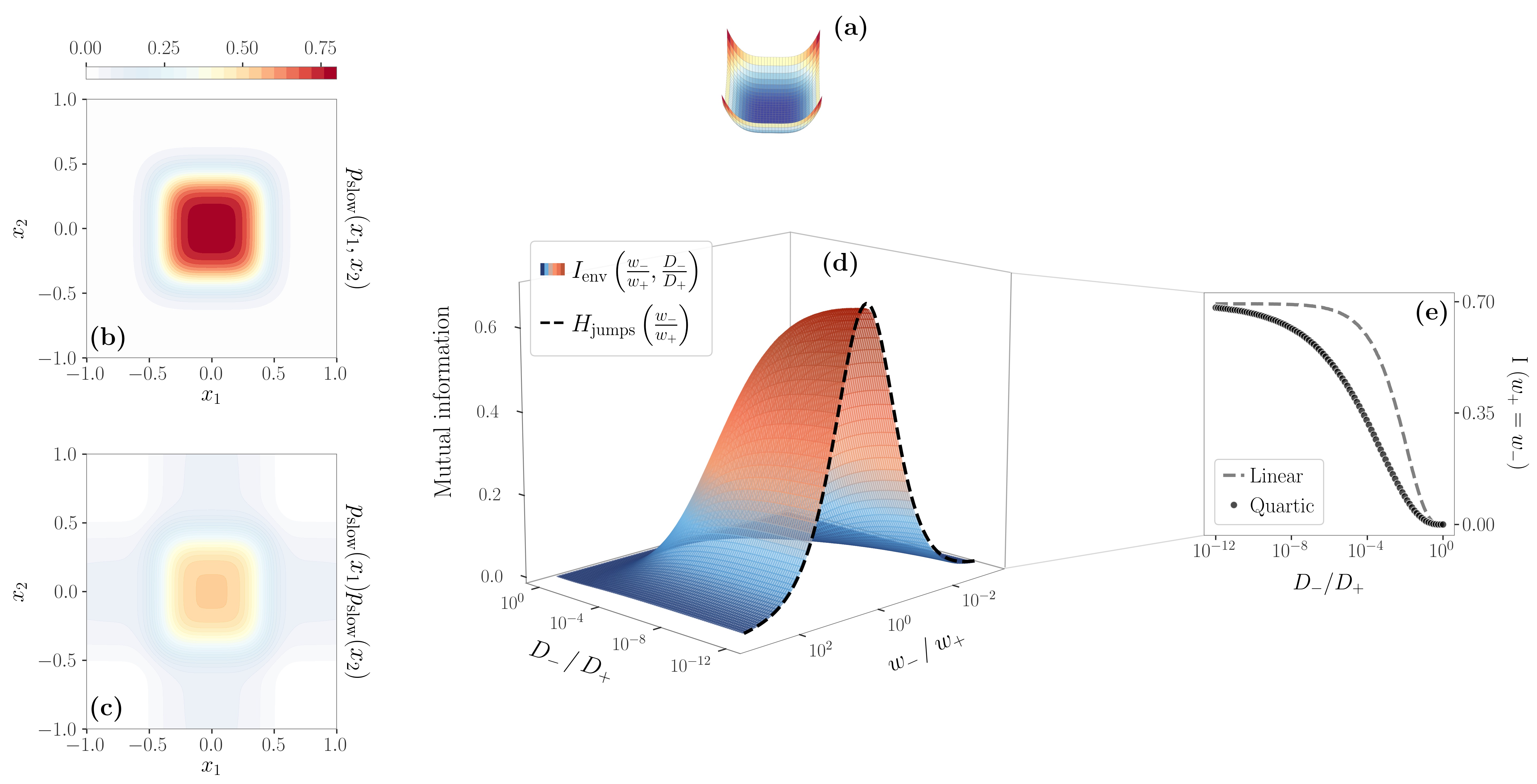}
    \caption{The environmental contribution to the mutual information as a function of $D_-/D_+$ and $w_-/w_+$ in a quartic potential and in the slow-jumps limit. For all plots, $\tau = 1$. (a) The quartic potential considered here. (b-c) Contour plots of the joint probability distribution and its factorization, respectively, for $D_+ = 10$, $D_- = 10^{-2}$, $w_- = w_+$. Notice that the marginalized probability has much longer tails along the axis $x = 0$ and $y = 0$. (d) The colored surface is the result of a Monte Carlo integration with importance sampling of the mutual information. In the $D_-/D_+ \to 0$ limit, $I_\mathrm{env}$ approaches $H_\mathrm{jumps}$, the black dashed line, which is also its maximum value. (e) Compared with the linear case, the non-linear relaxation reflects into a considerably slower convergence towards $H_{\rm jumps}$ of the mutual information.}
    \label{fig:mutual_quartic}
\end{figure*}

Let us start with a non-linear, yet non-interacting, case. Each particle diffuses in the $1D$ potential
\begin{equation}
    \label{eqn:quartic_potential}
    U(x_\mu) = \frac{x_\mu^4}{4 \tau},
\end{equation}
depicted in Fig.~\ref{fig:mutual_quartic}a. Hence, the drift term in Eq.~\eqref{eqn:fokker_planck} is simply given by the potential gradient $F_\mu(x_\mu) = -\partial_\mu U(x_\mu) = -x_\mu^3/\tau$, resulting in a non-linear relaxation. In the slow-jumps limit, this term leads to the following mixture components:
\begin{equation}
\label{eqn:quartic_mixture_joint}
    P_i^{\rm st}(x_1, x_2) = \sqrt{\frac{D_i \tau \pi^2}{2}}\frac{\Gamma\left(\frac{3}{4}\right)}{\Gamma\left(\frac{1}{4}\right)} \exp\left(-\frac{x_1^4+x_2^4}{4 D_i \tau}\right)
\end{equation}
which are the solutions of Eq.~\eqref{eqn:mixture_components}. The corresponding marginal distributions are 
\begin{equation}
\label{eqn:quartic_mixture_marg}
    P_i^{\rm st}(x_\mu) = \frac{\Gamma\left(\frac{3}{4}\right)}{\pi (D_i \tau)^{1/4}} e^{-\frac{x_\mu^4}{4 D_i \tau}}
\end{equation}
and, clearly, $P_i^{\rm st}(x_1,x_2) = P_i^{\rm st}(x_1)P_i^{\rm st}(x_2)$ since the particles are not interacting. The joint mixture distribution, Eq.~\eqref{eqn:solution_slow_jumps}, and its factorization,
\begin{equation}
     \prod_{\mu=1}^2 p_{\rm slow}(x_\mu) = \prod_{\mu=1}^2 \Bigg( \sum_{i = \{+,-\}} \left[\pi_i P_i^{\rm st}(x_\mu)\right] \Bigg),
\end{equation}
are plotted in Fig.~\ref{fig:mutual_quartic}b-c, and the mutual information in the slow-jumps limit corresponds to an information metric quantifying their distance. Notably, the effects of the environment in the joint distribution, with respect to the factorized probability distribution, reflect into a suppression of the tails along the axes.

As for the linear case \cite{nicoletti2021mutual}, we are able to compute the pairwise distance bounds \cite{Kolchinsky2017} on the mutual information analytically, starting from Eqs.~\eqref{eqn:quartic_mixture_joint}-\eqref{eqn:quartic_mixture_marg}. In particular, we find that the lower bound obeys
\begin{align}
\label{eqn:quartic_low_bound}
    I_\text{env}^{\rm{low}}\left(\frac{D_-}{D_+}, \frac{w_-}{w_+} \right) = & -\pi_+ \log\frac{\left(\pi_+ + h^1_{\rm C}\left(\frac{D_-}{D_+}\right)\pi_-\right)^2}{\pi_+ + h^2_{\rm D}\left(\frac{D_-}{D_+}\right)\pi_-} + \nonumber \\
    & -\pi_- \log\frac{\left(h^1_{\rm C}\left(\frac{D_+}{D_-}\right)\pi_+ + \pi_-\right)^2}{h^2_{\rm D}\left(\frac{D_+}{D_-}\right)\pi_+ + \pi_-}
\end{align}
where $h^1_{\rm C}(x) = (4 x)^{1/8}/(1+x)^{1/4}$ and $h_{\rm D}^1(x) = (x -1  - \log x)/4$ are, respectively, the $1/2$-Chernoff and Kullback-Leibler divergence between the $-$ and $+$ components of the marginal distribution. The divergences between the analogous components of the joint distribution are, respectively, $h^2_{\rm C}(x) = 2 h^1_{\rm C}(x)$ and $h^2_{\rm D}(x) = 2 h^1_{\rm D}(x)$. The upper bound $I_\text{env}^{\rm{up}}$ is identical to Eq.~\eqref{eqn:quartic_low_bound}, with the exchange $h_C \leftrightarrow h_D$.

Crucially, these bounds converge to the same limits of the linear case, namely,
\begin{align}
\label{eqn:mutual_extrinsic_limits}
    I_\text{env}\left(\frac{D_-}{D_+}, \frac{w_-}{w_+}\right) =
    \begin{cases}
        H_{\rm jumps} &\, \text{if} \quad D_-/D_+ \ll 1 \\
        \,0 &\, \text{if} \quad D_-/D_+ \approx 1
    \end{cases},
\end{align}
but their convergence rate is slower than the one obtained in the linear regime. This is perhaps unsurprising, since the non-linear relaxation increases the typical auto-correlation timescale and thus reduces the impact of environmental changes. In Fig.~\ref{fig:mutual_quartic}d, we show the mutual information in this slow-jumps limit, computed via importance sampling \cite{landau2021guide}. In particular, we sample the components of the joint distribution, defined in Eq.~\eqref{eqn:mixture_components}, starting from the potential in Eq.~\eqref{eqn:quartic_potential} via Hamiltonian Montecarlo \cite{neal2011mcmc, betancourt2017conceptual}. Then, each component is weighted according to the stationary distribution of the environment, for any given $w_-/w_+$, to obtain samples of Eq.~\eqref{eqn:solution_slow_jumps}. In Fig.~\ref{fig:mutual_quartic}e we see that, at a given value of $D_-/D_+$, the mutual information due to the environment is typically smaller than the case of linear relaxation.

It is possible to show that, for any potential of the form $U(x_\mu) \propto x_\mu^{2n}$, with $n$ positive integer, the bounds in Eq.~\eqref{eqn:quartic_low_bound} always saturate to $H_{\rm jump}$ when $D_-/D_+ \to 0$ and vanish when $D_- \to D_+$. This result, consistently with the one presented for linear interactions in \cite{nicoletti2021mutual}, remarks that, when the variability of the environment is maximal, any two non-interacting degrees of freedom share the information contained into the Shannon entropy associated with the external jump process, $H_{\rm jumps}$. Importantly, in these non-interacting cases, the only dimensionless parameters we can build are $w_-/w_+$, which determines the persistence of the two environmental states, and $D_-/D_+$, describing how similar the environmental states are. Although the probability distributions in Eq.~\eqref{eqn:quartic_mixture_marg} and Eq.~\eqref{eqn:quartic_mixture_joint} do not depend only on such combinations, we expect the mutual information to do so (see also \cite{nicoletti2021mutual}).
%Mutual information in a quartic potential.

\section{Mutual information in non-linear potentials}

We now consider non-linear interactions between the two particles. To keep things analytically tractable, we assume that the drift term in Eq.~\eqref{eqn:fokker_planck} can be written as the gradient of a potential of the form
\begin{equation}
\label{eqn:interacting_potentials}
    V(x_1, x_2) = \sum_\mu U(x_\mu) + V_{\rm int} (x_1, x_2)
\end{equation}
so that the solution of Eq.~\eqref{eqn:mixture_components} is given by the Boltzmann-like form $P_i^{\rm st}(x_1, x_2) \propto {\rm exp}\left[-V(x_1,x_2)/D_i\right]$. Hence, we are focusing on equilibrium systems with non-linear relaxation and non-linear interactions.

For convenience, we write the mutual information of the overall system as follows:
\begin{gather}
    \label{eqn:mutual_decomposition}
    I\left(\frac{w_-}{w_+}, \frac{D_-}{D_+}, \{\psi\}\right) = \\
    = I_{\rm env}\left(\frac{w_-}{w_+}, \frac{D_-}{D_+}\right) + I^{\rm int}\left(\{\psi\} \right) + \Xi \left(\frac{w_-}{w_+}, \frac{D_-}{D_+}, \{\psi\} \right) \nonumber
\end{gather}
where $\{\psi\}$ is the set of dimensionless parameters associated with the interactions. The first term, $I_{\rm env}$, is the mutual information stemming from the shared environment. It can only depend on the environmental dimensionless parameters, $D_-/D_+$ and $w_-/w_+$. The second term, $I^{\rm int}$, is instead the mutual information associated with the joint distribution $P_i^{\rm st}(x_1, x_2)$. This is the term that we would expect in the absence of the environment, and thus can only depend on $\{\psi\}$. Finally, $\Xi$ quantifies the contribution to the mutual information due to the presence of both the environment and the interactions at once. In general, this is not a mutual information, i.e., it needs not to be positive, and may depend on all dimensionless parameters. For these reasons, we name this term as \emph{information interference}. 

Unless otherwise specified, the mutual information in the slow-jumps regime is obtained as outlined before. We employ Hamiltonian Montecarlo to sample the joint distribution associated with the potential in Eq.~\eqref{eqn:interacting_potentials}, and we weight these samples according to the corresponding mixture distribution, Eq.~\eqref{eqn:solution_slow_jumps}. Then, the mutual information integral is evaluated via importance sampling. Crucially, importance sampling requires the knowledge of the analytical expressions of both the joint and the marginal mixture components \cite{landau2021guide}, which we need to compute for every choice of the potential.

\begin{figure*}[t]
    \centering
    \includegraphics[width=0.98\textwidth]{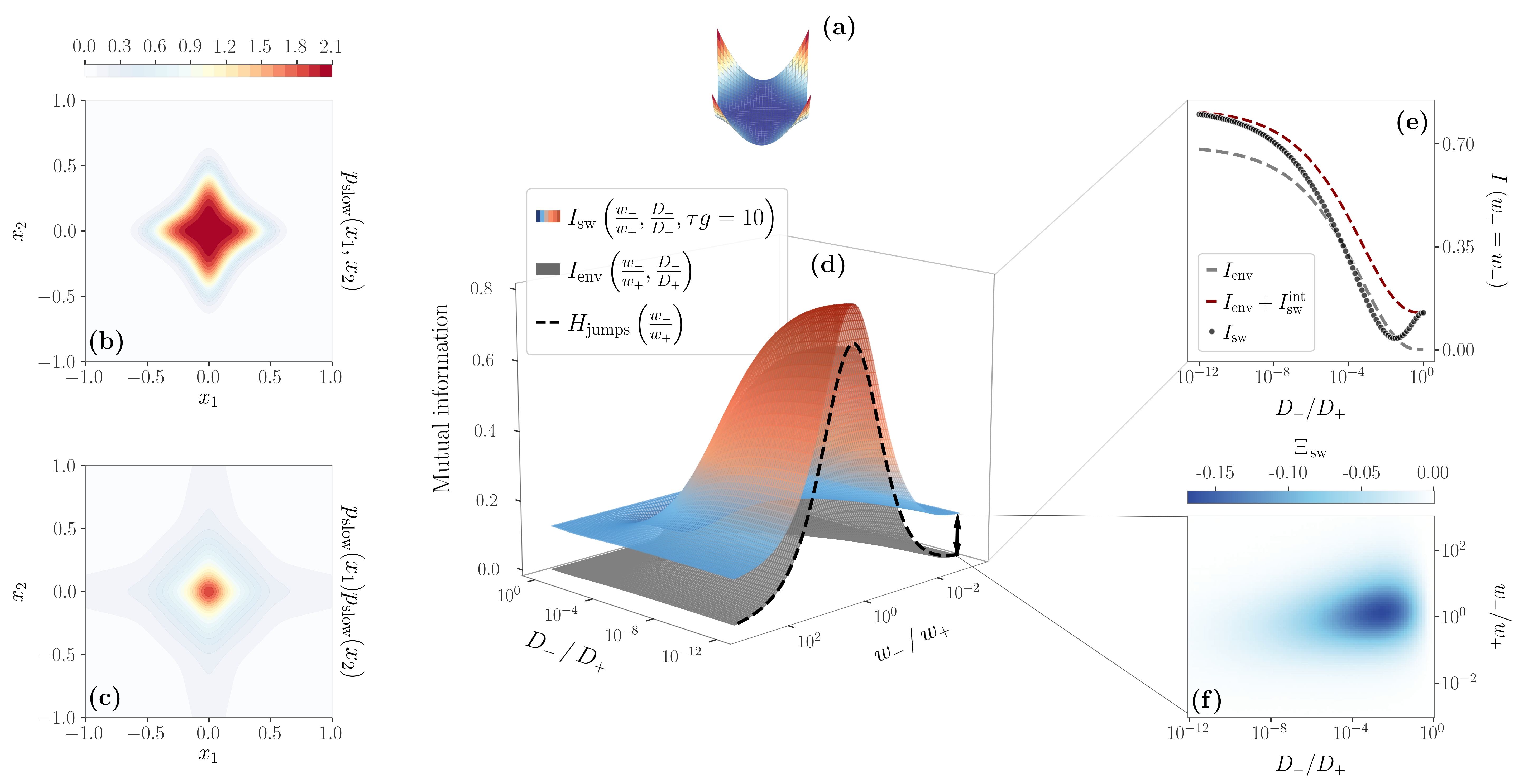}
    \caption{The total mutual information $I_{\rm sw}$ in the single-well case, as a function of $D_-/D_+$ and $w_-/w_+$, in the slow-jumps limit. For all plots, $\tau g = 10$. (a) The single-well potential considered here. (b-c) Contour plots of the joint probability distribution and its factorization, respectively, for $D_+ = 10$, $D_- = 10^{-2}$, $w_- = w_+$. The marginalized probability has much longer tails along the axis $x= 0$ and $y= 0$, which tend to be suppressed by the environment. However, these tails are still present in the joint probability as a consequence of the interactions. (d) The colored surface is the result of a Monte Carlo integration with importance sampling of $I_{\rm sw}$, whereas the gray surface represent the environmental contribution alone $I_{\rm env}$. (e) $I_{\rm sw}$ (black dots) can be smaller than $I_{\rm env}$ (gray dashed line) and, in general, it is lower than the sum of $I^{\rm int}_{\rm sw}$ and $I_{\rm env}$ (red dashed line). (f) In fact, the term $\Xi_{\rm sw}$ is always negative, showing that the effects of the environment and of the interactions are reciprocally masked at low-enough values of $D_-/D_+$. For $D_-/D_+ \to 0, 1$ we find $\Xi_{\rm sw} = 0$, hence the environmental and the interactions contributions are disentangled.}
    \label{fig:mutual_single_well}
\end{figure*}

\subsection{Destructive information interference}
We first study the case
\begin{equation}
    \label{eqn:single_well_potential}
    V(x_1, x_2) = \frac{x_1^4 + x_2^4}{4\tau} - g \frac{x_1^2 x_2^2}{2} := V_{\rm sw} (x_1, x_2)
\end{equation}
where for stability $g > 0$. This single-well potential, depicted in Fig.~\ref{fig:mutual_single_well}a, has one stable minima at $(x_1, x_2) = (0,0)$. In the slow-jumps limit, the mixture components of the joint distributions follows a Boltzmann-like distribution
\begin{equation}
    \label{eqn:single_well_joint}
    P_i^{\rm st}(x_1, x_2) = \frac{1}{\mathcal{N}_{\rm sw}}e^{-V_{\rm sw}(x_1, x_2)/D_i}
\end{equation}
where the normalization $\mathcal{N}_{\rm sw}$ can be computed analytically. The corresponding marginal components are
\begin{equation}
    \label{eqn:single_well_marginal}
    P_i^{\rm st}(x_\mu) = \sqrt{\frac{g x_\mu^2 \tau}{2 \mathcal{N}_{\rm sw}^2}} K_\frac{1}{4}\left(\frac{g^2x_\mu^4\tau}{8 D_i}\right)e^{(-2+g^2\tau^2)\frac{x_\mu^4}{8 D_i \tau}}
\end{equation}
where $K_{n}(x)$ is the modified Bessel function of the second kind. With this choice of the potential, the only dimensionless parameters appearing in the mutual information are $D_-/D_+$, $w_-/w_+$, and $g \tau$. The first two belong to the environment, while the last one is the sole quantity characterizing the interactions. Notice that we only have to inspect the dynamics for a fixed environment to determine the dimensionless relevant quantities, and add $w_-/w_+$ that modulates the mixture in the slow-jumps limit. In other words, and as for the linear case, the mutual information of the joint distribution, $I_{\rm sw}$, cannot depend separately on $D_-$ and $D_+$ thus being independent of the environmental state.

\begin{figure*}[t]
    \centering
    \includegraphics[width=0.98\textwidth]{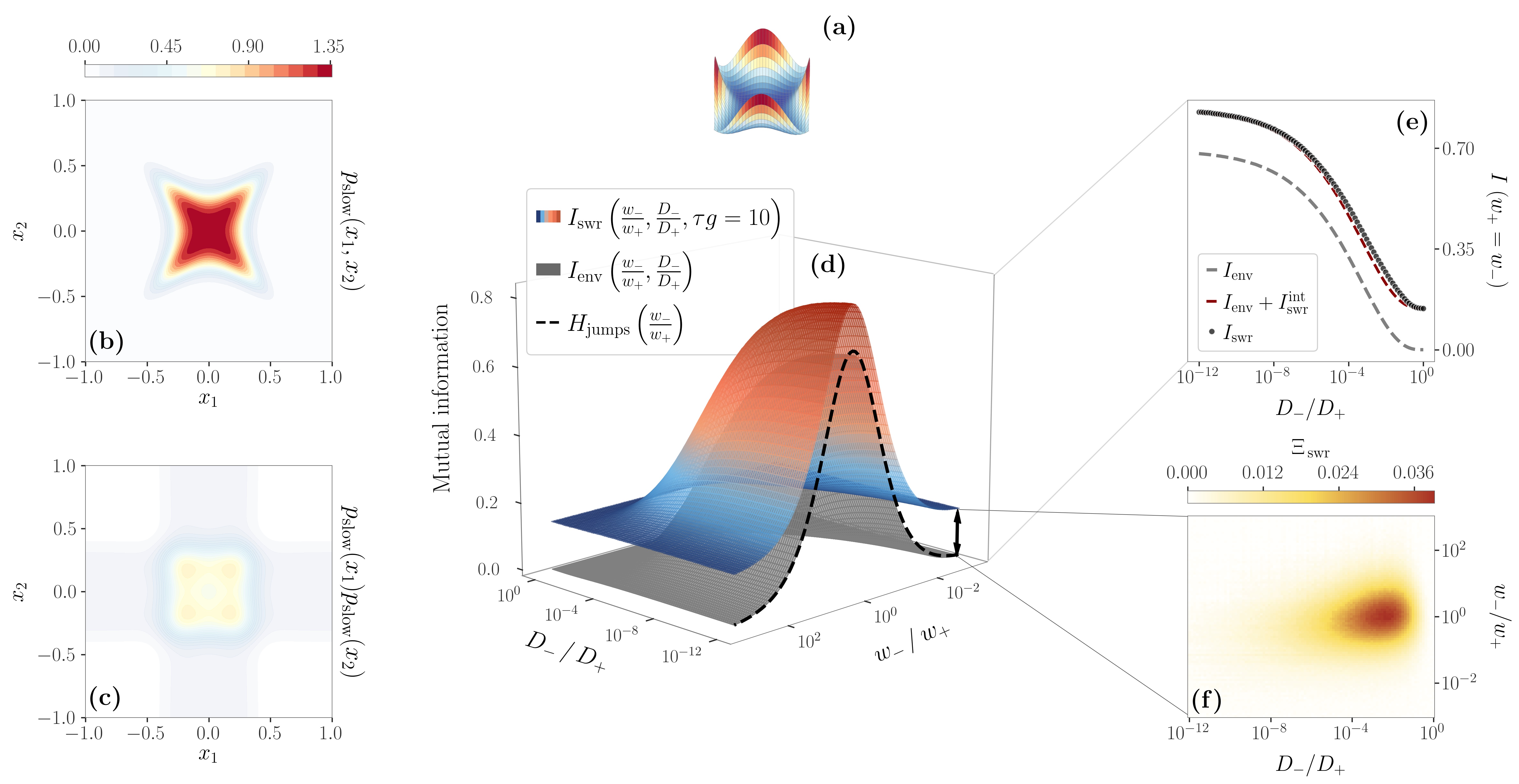}
    \caption{The total mutual information $I_{\rm swr}$ in the rotated single-well case, as a function of $D_-/D_+$ and $w_-/w_+$, in the slow-jumps limit. For all plots, $\tau g= 10$. (a) The rotated single-well potential considered here. (b-c) Contour plots of the joint probability distribution and its factorization, respectively, for $D_+ = 10$, $D_- = 10^{-2}$, $w_- = w_+$. The marginalized probability has much longer tails along the axis $x= 0$ and $y= 0$, which are suppressed in the joint probability as a consequence of the environment. Contrary to the single-well case, the interactions here trigger the presence of tails along the bisectors of the $(x,y)$ plane. (d) The colored surface is the result of a Monte Carlo integration with importance sampling of $I_{\rm swr}$, whereas the gray surface represent the environmental contribution alone $I_{\rm env}$. (e) $I_{\rm swr}$ (black dots) is always greater than $I_{\rm env}$ (gray dashed line) and, in general, it is greater than the sum of $I^{\rm int}_{\rm swr}$ and $I_{\rm env}$ (red dashed line). (f) The term $\Xi_{\rm swr}$ is always positive, and in particular it is different from zero at high enough values of $D_-/D_+$. For $D_-/D_+ \to 0, 1$ we end up with  $\Xi_{\rm swr} = 0$ as expected.}
    \label{fig:mutual_single_well_rotated}
\end{figure*}

Consequently, we write $I_{\rm sw}$ as
\begin{gather}
    \label{eqn:single_well_mutual_decomposition}
    I_{\rm sw}\left(\frac{w_-}{w_+}, \frac{D_-}{D_+}, \tau g \right) = \\
    = I_{\rm env}\left(\frac{w_-}{w_+}, \frac{D_-}{D_+}\right) + I_{\rm sw}^{\rm int}\left(\tau g \right) + \Xi_{\rm sw} \left(\frac{w_-}{w_+}, \frac{D_-}{D_+}, \tau g \right) \nonumber.
\end{gather}
Let us investigate separately the impact of interactions and environmental changes in this example. First, the effect of the interactions in the joint probability distribution reflects into the appearance of tails along the axes $x_1 = 0$ and $x_2 = 0$ (see Fig.~\ref{fig:mutual_single_well}b-c). The higher is $g$, the longer the tails. Conversely, the environment affects the joint distribution by suppressing such tails as the difference between $D_-$ and $D_+$ becomes more pronounced. Since these two terms operate in opposite ways, the mutual information takes contrasting contributions. As a consequence of this interplay between changing environment and non-linear interactions, in Fig.~\ref{fig:mutual_single_well}d-e-f, we see the mutual information of the overall system, $I_{\rm sw}$, is always smaller than the sum of $I_{\rm env}$ and $I_{\rm sw}^{\rm int}$, and can also be smaller than $I_{\rm env}$ for some values of $(w_-/w_+,D_-/D_+)$. This means that $\Xi \leq 0$ in the entire space, and $\Xi < -I_{\rm sw}^{\rm int}$ in some regions of the parameter space (see Fig.~\ref{fig:mutual_single_well}f). Naively speaking, non-linear interactions can mask environmental information, by counteracting the dependency induced by a switching environment and effectively reducing the information that $x_1$ and $x_2$ share. We name this phenomenon \textit{destructive information interference}.

However, the limiting behaviors of $I_{\rm sw}$ can be understood as discussed in \cite{nicoletti2021mutual}, and the disentangling is possible provided some knowledge of the environmental states in these regimes. Indeed, when $D_-/D_+ \to 1$, the only contribution to the mutual information comes from the interactions alone, $I_{\rm sw}^{\rm int}(\tau g)$. Similarly, in the opposite limit $D_-/D_+ \to 0$, we find that the two contributions to the mutual information are exactly disentangled, i.e., 
\begin{equation}
    \label{eqn:single_well_limits}
    I_{\rm sw}\left(\frac{w_-}{w_+}, \frac{D_-}{D_+}, \tau g \right) = 
    \begin{cases}
        H_{\rm jumps} + I_{\rm sw}^{\rm int}(\tau g) & \text{if} \,\,\, \frac{D_-}{D_+} \ll 1 \\
        \,I_{\rm sw}^{\rm int}(\tau g) & \text{if} \,\,\, \frac{D_-}{D_+} \approx 1
    \end{cases}
\end{equation}
that means that in both limits $\Xi \to 0$. In Fig.~\ref{fig:mutual_single_well}e, we compare the fully disentangled form (in red) with the mutual information obtained with non-linear relaxation and interactions as in Eq.~\eqref{eqn:single_well_potential}.

%Qualitatively, the interacting term in Eq.~\eqref{eqn:single_well_potential} counterbalances the environmental effects, which tend instead to suppress such tails, as shown in the previous section. In Fig.~\ref{fig:mutual_single_well}b-c we see these two contributions at play.

%Mutual information in the single-well (environment and interactions mask one another). Mutual information in the rotated single-well case (environment and interactions act almost separately, and can enhance one another). Mutual information in the double-well case (large enhancing, different adimensional parameters).

\subsection{Constructive information interference}
In the previous section, we argued that  the destructive information interference stems from the fact that interactions and environment operate on the same axes in opposite ways. Indeed, we now show that a rotation of the interaction term in Eq.~\eqref{eqn:single_well_potential} of an angle $\pi/4$ generates instead a cooperation of the two terms that can boost the overall mutual information. In analogy with the previous case, this features is named \textit{constructive information interference}.

Thus, the potential governing the system, shown in Fig.~\ref{fig:mutual_single_well_rotated}a, is:
\begin{equation}
    \label{eqn:rotated_single_well_potential}
    V_{\rm swr}(x_1, x_2) = \frac{x_1^4 + x_2^4}{4\tau} - g \frac{||R_{\pi/4}(x, y)||^2}{2}
\end{equation}
where $R_\theta$ is the rotation matrix of angle $\theta$, $||\cdot||^2$ is the $L_2$ norm and $g>0$.

In this scenario, the joint and marginal mixture components in the slow-jumps limit can be again found analytically, and are given by
\begin{equation}
    \label{eqn:rotated_single_well_joint}
    P_i^{\rm st}(x_1, x_2) = \frac{1}{\mathcal{N}_{\rm swr}}e^{-V_{\rm swr}(x_1, x_2)/D_i}
\end{equation}
and 
\begin{align}
    \label{eqn:rotated_single_well_marginal}
    P_i^{\rm st}(x_\mu) = & \sqrt{\frac{g\pi^2}{2 \alpha}} \frac{|x_\mu| \left[I_{-\frac{1}{4}}\left(\beta_i x_\mu^4\right) + I_{\frac{1}{4}}\left(\beta_i x_\mu^4\right)\right]}{2\mathcal{N}_{\rm swr}} \times \nonumber\\
    & \times e^{\frac{x_\mu^4(g^2 - 128 \alpha)}{128\alpha D_i}}
\end{align}
where $I_n(x)$ is the modified Bessel function of the first kind, $\alpha = \tau^{-1}/4 + g/8$ and $\beta_i = g^2/(128 \alpha D_i)$. 

As in the previous case, the dimensionless parameters appearing in the mutual information are $w_-/w_+$, $D_-/D_+$, and $\tau g$. Hence, the interference term in Eq.~\eqref{eqn:mutual_decomposition}, $\Xi_{\rm swr}$, will again depend on all of them. However, in this case, the role of the interactions is to introduce tails along the bisectors of the $(x_1, x_2)$ plane, whereas the environment keeps acting on the $x_1 = 0$ and $x_2 = 0$ axes. Hence, non-linear interactions do not counteract the dependency induced by the environment. As a consequence, as shown in Fig.~\ref{fig:mutual_single_well_rotated}d-e, the mutual information of the overall system is very close to the sum of the environmental and the interaction terms, i.e., $\Xi_{\rm swr} \approx 0$. Moreover, there is a region in the parameter space in which $\Xi_{\rm swr} > 0$, meaning that $x_1$ and $x_2$ share more information than the one coming from the changing environment and their sheer couplings.

We remark that the limiting behaviors of the mutual information exhibit an exact disentangling, as before. Thus, when $D_-/D_+ \to 0$, $\Xi_{\rm swr} \to 0$ and $I_{\rm env} \to H_{\rm jumps}$, while for $D_- \to D_+$ only $I_{\rm swr}^{\rm int}$ survives.

%As we can see in Fig.~\ref{fig:mutual_single_well_rotated}b-c, the effect of the interactions in these distribution is the presence of longer tails along the bisectors of the $(x_1, x_2)$ plane rather than the axes. As a consequence, the interactions do not counteract the dependency induced by the environment as before.

%Once more, the only dimensionless parameters in the slow-jumps limit are $D_-/D_+$, $w_-/w_+$ and $\tau g$ - hence, the mutual information associated to Eq.~\eqref{eqn:rotated_single_well_joint} is a solely a function of the interactions $I_{\rm swr}^{\rm int}(\tau g)$. In Figures \ref{fig:mutual_single_well_rotated}d-e-f we plot the overall mutual information, $I_{\rm swr}$m as a function of the first two adimensional parameters, and we compare it with $I_{\rm env}$. 
%As expected, we find that $I_{\rm swr}$ is always greater than $I_{\rm env}$ - i.e., no masking of the mutual information occurs. Yet, for some values of $(w_-/w_+, D_-/D_+)$, we have $I_{\rm swr} > I_{\rm env} + I_{\rm swr}^{\rm int}$. Therefore, the opposite effect is taking place - the interactions are now enhancing the information coming from the environment.

\begin{figure*}[t]
    \centering
    \includegraphics[width=0.98\textwidth]{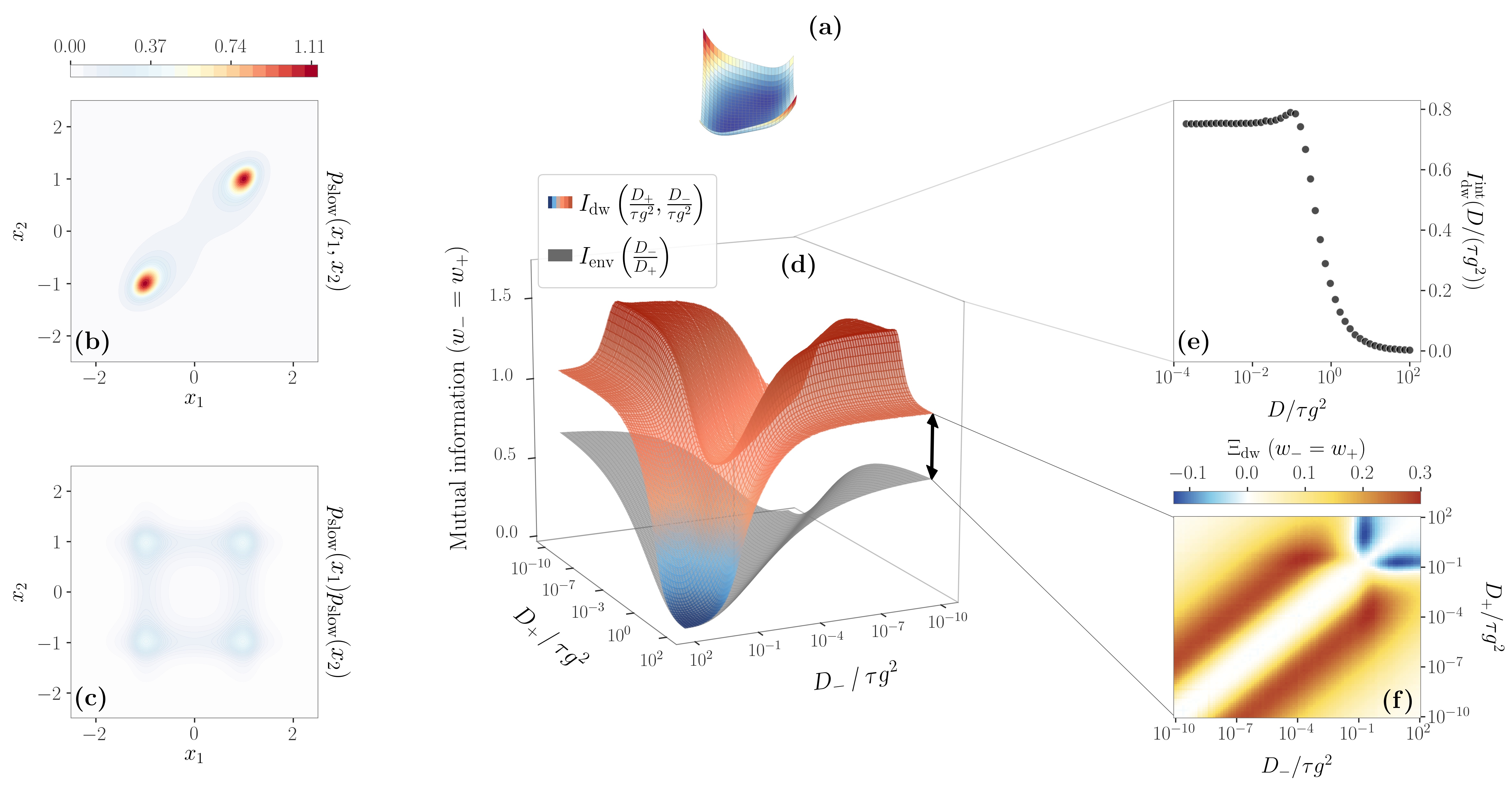}
    \caption{The total mutual information $I_{\rm dw}$ in the double-well case, as a function of the adimensional parameters $D_\pm/(\tau g^2)$, in the slow-jumps limit. For all plots, $w_-/w_+ = 1$, $\tau = g = 1$. (a) The double-well potential considered here. (b-c) Contour plots of the joint probability distribution and its factorization, respectively, for $D_+ = 10$, $D_- = 10^{-2}$. The joint probability has two peaks, corresponding to the two minima of the potential. On the contrary, the marginalized probability is markedly different, with four peaks. (d) The colored surface is the result of a Monte Carlo integration with importance sampling of $I_{\rm dw}$, whereas the gray surface represent the environmental contribution alone $I_{\rm env}$, in the plane $(D_-/\tau g^2, D_+/\tau g^2)$. (e) The mutual information $I_{\rm dw}^{\rm int}$ of the interactions only. At large $D$, we expect the two minima to be less relevant, and indeed the mutual information vanishes. At small $D$, instead, $I_{\rm dw}^{\rm int}$ is markedly different from zero since the particles are typically trapped in one of the two minima. At intermediate values the mutual information peaks due to an interplay between trapping and diffusion. (e) The term $\Xi_{\rm dw}$ can be either positive and negative, meaning that, at different values of $D_\pm/(\tau g^2)$, we find both constructive and destructive interference. Clearly, when $D_- \approx D_+$, we have $\Xi_{\rm dw} \approx 0$.}
    \label{fig:mutual_double_well}
\end{figure*}

\subsection{Information peaks in bistable systems}
As a last example, here we consider the slightly more complex case of a bistable system. In particular, we have the following potential:
\begin{equation}
    \label{eqn:double_well_potential}
    V_{\rm dw}(x_1, x_2) = \frac{x_1^4 + x_2^4}{4\tau} - g x y
\end{equation}
where $g>0$. This potential, depicted in Fig.~\ref{fig:mutual_double_well}a, has two stable minima at $(x_1,x_2) = (\pm \sqrt{g\tau}, \pm \sqrt{g\tau})$. The joint and the marginal mixture components in the slow-jumps limit are:
\begin{equation}
    \label{eqn:double_well_joint}
    P_i^{\rm st}(x_1, x_2) = \frac{1}{\mathcal{N}_{\rm dw}}e^{-V_{\rm dw}(x_1, x_2)/D_i}
\end{equation}
and 
\begin{align}
    \label{eqn:double_well_marginal}
    P_i^{\rm st}(x_\mu) = & \frac{e^{-\frac{x_\mu^4}{4 D_i \tau}}}{\sqrt{2}D_i^2 \mathcal{N}_\mathrm{dw}} \biggl[(D_i^9\tau)^\frac{1}{4} \Gamma\left(\frac{1}{4}\right) {_0}F_2\left(\frac{1}{2},\frac{3}{4}; \alpha_i x_\mu^4\right) + \nonumber\\
    & g^2x_\mu^2(D_i\tau)^\frac{3}{4} \Gamma\left(\frac{3}{4}\right) {_0}F_2\left(\frac{5}{4},\frac{3}{2}; \alpha_i x_\mu^4\right)\biggl]
\end{align}
where ${_p}F_q(a_1, \dots, a_p;b_1, \dots, b_q;x)$ is the generalized hypergeometric function, and $\alpha_i = g^4 \tau/(64D_i^3)$. As we can see in Fig.~\ref{fig:mutual_double_well}b-c, the joint probability distribution has two peaks corresponding to the two minima of the potential, whereas the factorized distribution presents four peaks with connections among them that reflect the influence of a switching environment.

The first crucial difference between this case and the previous ones is that the dimensionless parameters appearing in the mutual information mix environmental and interaction features. Indeed, they are $w_-/w_+$, $D_-/\tau g^2$, and $D_+/\tau g^2$, with now $D_-/D_+$ resulting from a combination of the last two parameters. This immediately informs us that the mutual information of the overall system, $I_{\rm dw}$, reported in Fig.~\ref{fig:mutual_double_well}d, will depend separately on the environmental states, $D_-$ and $D_+$. Therefore, we write Eq.~\eqref{eqn:mutual_decomposition} as follows:
\begin{gather}
    I_{\rm dw}\left(\frac{w_-}{w_+}, \frac{D_-}{\tau g^2}, \frac{D_+}{\tau g^2}\right) = I_{\rm env}\left(\frac{w_-}{w_+}, \frac{D_-}{D_+}\right) + \nonumber \\
    + \sum_i \pi_i I_{\rm dw}^{\rm int}\left(\frac{D_i}{\tau g^2}\right) + \Xi_{\rm dw}\left(\frac{w_-}{w_+}, \frac{D_-}{\tau g^2}, \frac{D_+}{\tau g^2}\right)
\end{gather}
where $\pi_i$ depends solely on $w_-/w_+$.

The dependence on the diffusion coefficient of the interaction term, $I_{\rm dw}^{\rm int}$, shown in Fig.~\ref{fig:mutual_double_well}e, can be explained on an intuitive basis. Indeed, the distance between $P_i^{\rm st}(x_1, x_2)$ and its factorization receives the most contributions from the fact that the latter has four peaks, due to the implicit assumption of independence between $x_1$ and $x_2$. However, when $D$ is large, the system can easily escape the potential minima, and thus they will not contribute to $I_{\rm dw}^{\rm int}$, which vanishes as $D$ grows.

Conversely, small values of $D$ weight more the potential minima, since the system is substantially trapped in them. In this limit, $I_{\rm dw}^{\rm int}$ converges to a non-zero value due to the fact that only two of the peaks of the factorized distribution are present in the joint distribution. Finally, we observe an emerging peak of $I_{\rm dw}^{\rm int}$ at a finite value of $D$. This optimal diffusion naively allows the system to explore both minima from time to time, still being trapped for a consistent amount of time during each stochastic realization.

In Fig.~\ref{fig:mutual_double_well}f, we report $\Xi_{\rm dw}$ for the specific case $w_- = w_+$. All other choices do not qualitatively change the picture. In this scenario, $\Xi_{\rm dw}$ can be either positive and negative, exhibiting a non-trivial pattern of constructive and destructive information interference. This pattern, although hard to understand analytically, is intuitively a consequence of the system switching from a state in which it is trapped in one single minimum, to a state in which it can freely explore larger regions of the $(x_1,x_2)$ plane.

We also remark that in this case it is difficult to define the usual limiting behaviors of the mutual information of the overall system in which the disentangling is recovered. Indeed, $D_-/D_+$ is not the only relevant parameter of the system and the limit $D_-/D_+ \to 0$ is not particularly informative anymore.

\section{Mutual information in non-equilibrium environments}
\begin{figure}[t]
    \centering
    \includegraphics[width=0.98\columnwidth]{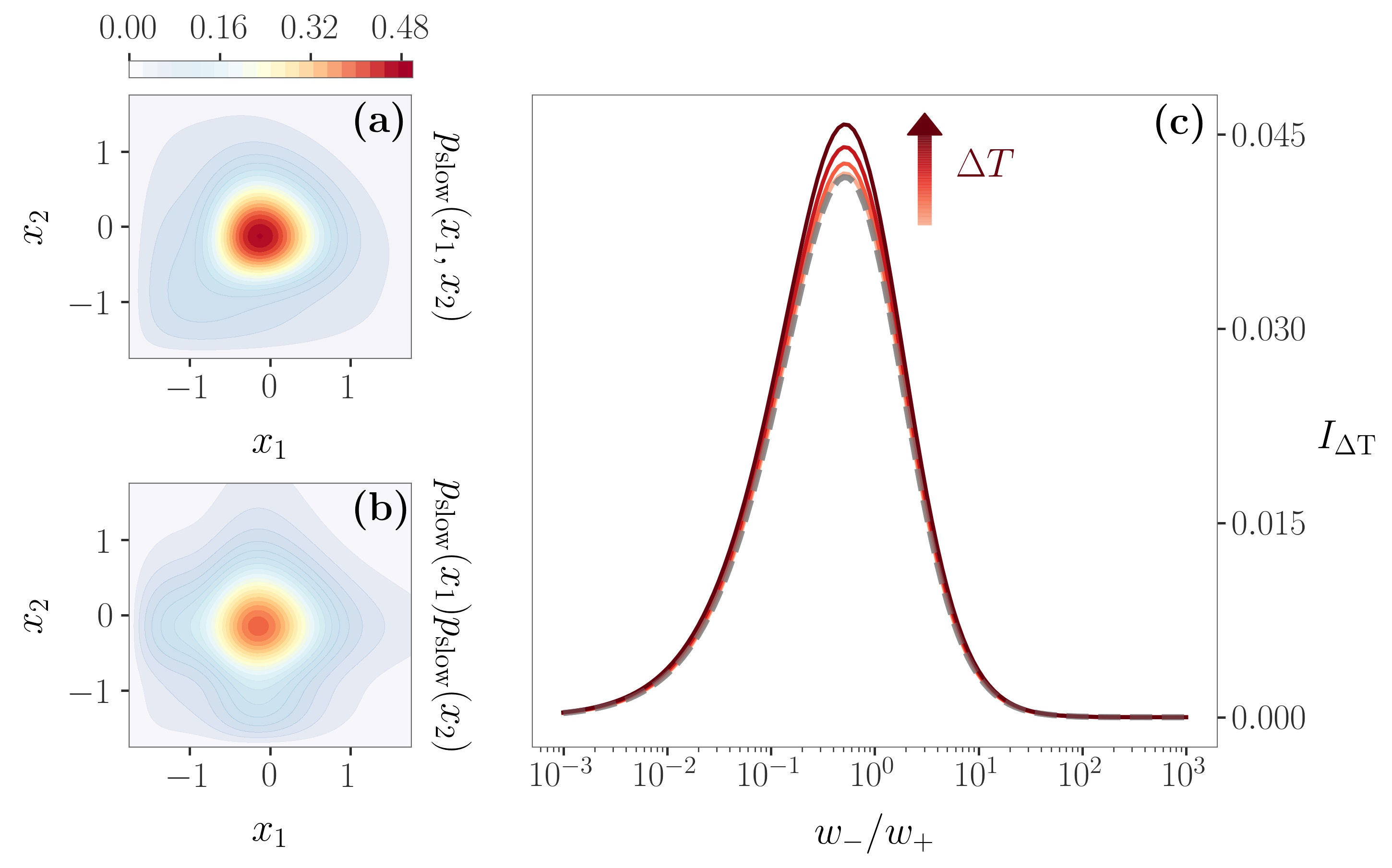}
    \caption{Mutual information in the presence of multiplicative noise. (a-b) The joint and the marginal probability distributions in the slow-jumps limit. (c) As we increase the temperature gradient $\Delta T$, the mutual information increases. Here, the different curves are for $\Delta T = \{0.15, 0.29, 0.43, 0.57\}$, $T_0 = \tau = 1$, $\gamma_+ = 1.5$ and $\gamma_- = 0.5$.}
    \label{fig:multiplicative}
\end{figure}
So far, we investigated systems coupled with an ever-changing environment that eventually relax to equilibrium. However, nature usually operates out-of-equilibrium, and most of the environments of biochemical, neural, and ecological systems are in non-equilibrium conditions. To study a minimal model encompassing this additional feature, we consider the following dynamics:
\begin{equation}
    \dot{x}_\mu = - \frac{1}{\tau} x_\mu + \gamma_{i(t)} \sqrt{2 T(x_\mu)} \xi_\mu
\end{equation}
with $i(t)$ is a realization of the stochastic process governing the environment, $\mu = 1, 2$, and $T(x_\mu) = T_0 + x_\mu \Delta T$ is a linear temperature gradient, for the sake of simplicity. This picture can capture the key features of a diffusing molecule that can live in two conformational states \cite{barducci2015non,gaspari2007aggregation,busiello2021dissipation,liang2021intrinsic}. Alternatively, it can be a simple way to describe proteins in an environment with patches of different density (e.g., liquid condensates \cite{weber2019physics,hyman2014liquid}) subject to an external gradient. Since diffusion and temperature are connected by the Einstein relation, $D(x) \propto \gamma T(x)$, here the environment may act as a modification of the viscosity, in the case of patches of different density, or the motility, when the switching describes two different conformational states. These diffusive properties are encoded into $\gamma_{i(t)}$ that can take two values, $\gamma_-$ and $\gamma_+$, replacing the role of $D_-$ and $D_+$ of the previous models.

The peculiarity of this model is the presence of a multiplicative noise proportional to $x_\mu$. Since there are no interactions, the joint component is just the product of the two mixture components. Here, $x_\mu \in [-T_0/\Delta T, +\infty]$, and $\Delta T < \sqrt{T_0/\gamma \tau}$ to ensure flat derivatives at the boundaries, so that no particles can escape the system. Hence, the mixture components read:
\begin{equation}
    P_i(x_\mu) = \mathcal{N} e^{-\frac{1}{\tau} \frac{x_\mu}{\gamma \Delta T}} \left( 1 + \frac{\Delta T}{T_0} x_\mu \right)^{\frac{1}{\tau}\frac{T_0/\Delta T}{\gamma \Delta T} - 1}
\end{equation}
In Fig.~\ref{fig:multiplicative}a-b, we respectively show the joint and factorized distributions of this system, that are both symmetric with respect to the bisector $x_1 = x_2$. In Fig.~\ref{fig:multiplicative}c, the mutual information in the slow-jumps limit is reported as a function of $w_-/w_+$ for increasing $\Delta T$. We can effectively conclude that the presence of a multiplicative noise increase the shared information between two (non-interacting) degrees of freedom.

Finally, as in the presence of a non multiplicative noise, the mutual information vanishes when $\gamma_- \to \gamma_+$, whereas it converges to $H_{\rm jumps}$ for $\gamma_-/\gamma_+ \to 0$, hence preserving the limiting behaviors that are crucial to perform an exact disentangling \cite{nicoletti2021mutual}.

\section{Mutual information in continuous environments}

Finally, we consider the case in which two particles are not interacting, but share the same continuous environment \cite{Chechkin2017,Wang2020}. To fix the ideas, let us consider the paradigmatic example of two Ornstein-Uhlenbeck processes,
\begin{equation}
\label{eqn:continuous_env_langevin}
    \begin{cases}
        \dot{x}_\mu = -x_\mu/\tau_X + \sqrt{2} D \,\xi_\mu \\
        \dot{D} = -D/\tau_D + \sqrt{2 \theta} \,\xi_D
    \end{cases}
\end{equation}
where the only adimensional parameter of the system is now $\frac{\tau_X}{\tau_D}$, which governs the time-scale separation of the two dynamics. Hence, contrary to the case of a discrete-state environment, we cannot define the separation between environmental states - previously quantified by $D_-/D_+$ - nor their relative persistence - which was given by $w_-/w_+$.

The corresponding stationary Fokker-Planck equation is given by
\begin{align}
    \label{eqn:continuous_env_FP}
    0 = & \sum_{\mu = 1}^2\left[\partial_\mu\left(\frac{x_\mu}{\tau_X} p(\vb{x}, D)\right) + D^2\partial_\mu^2 p(\vb{x}, D)\right] + \nonumber \\
    & + \partial_D \left[\frac{D}{\tau_D} p(\vb{x}, D)\right] + \theta \partial_D^2 p(\vb{x}, D)
\end{align}
and, as before, we are interested in the marginalization $p(\vb{x},t) = \int dD p(\vb{x}, D, t)$. Notably, if we explicitly marginalize Eq.~\eqref{eqn:continuous_env_FP}, at stationarity we obtain
\begin{align}
    \label{eqn:continuous_env_FP_marg}
    0 = & \sum_{\mu = 1}^2\left[\partial_\mu\left(\frac{x_\mu}{\tau_X} p(\vb{x})\right) + \partial_\mu^2\left(\hat{D}^2(\vb x) p(\vb{x})\right)\right] 
\end{align}
where $\hat{D}^2(\vb x) = \int dD\,D^2 p(D|\vb{x})$ is an effective spatial diffusion coefficient. Therefore, we can interpret the effective dependencies induced by the environment as arising from an inhomogeneous media, rather than associated with effective couplings between $x_1$ and $x_2$. It is also worth noting that, in principle, space-dependent diffusion coefficients, interpreted in the Ito sense, might always emerge from the variations of an external stochastic environment, which is also the sole responsible for a non-zero mutual information. This result might shed some light on the controversial topic about the Ito/Stratonovich dilemma in diffusing chemical systems. A similar perspective, where the internal states play an analogous role of a changing environment, is presented in \cite{liang2021intrinsic}.

\begin{figure}[t]
    \centering
    \includegraphics[width=0.98\columnwidth]{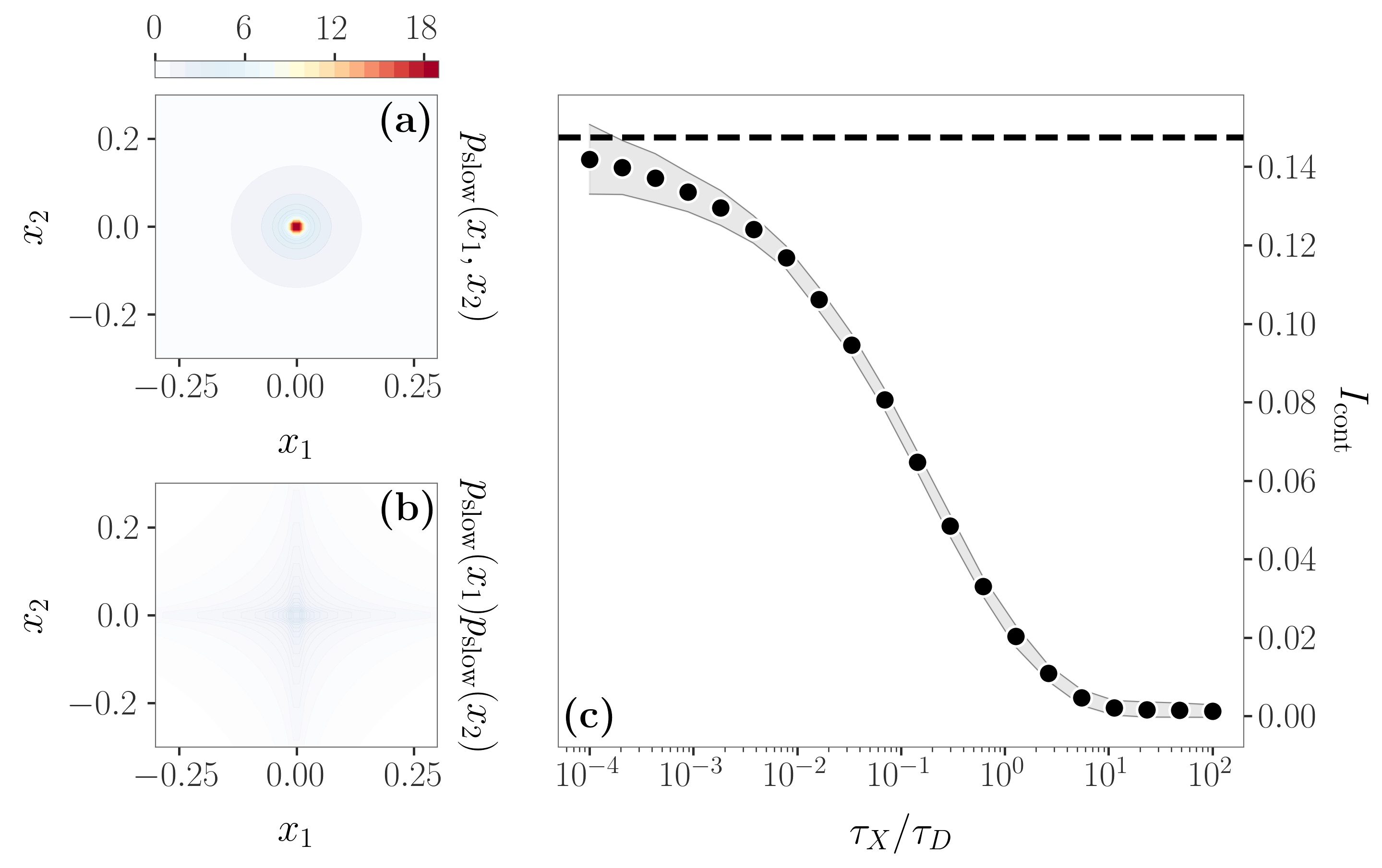}
    \caption{The mutual information associated to the Langevin equations in Eq.~\eqref{eqn:continuous_env_langevin}. (a-b) Plots of the joint and the factorized distribution in the limit of a much slower environment, $\tau_X/\tau_D \ll 1$. (c) Mutual information at different values of $\tau_X/\tau_D$ estimated through a k-nearest neighbors estimator. As expected, in the limit $\tau_X/\tau_D \ll 1$ the mutual information converges to Eq.~\eqref{eqn:continuous_env_integral}, whereas it vanishes in the opposite limit.}
    \label{fig:continuous_env}
\end{figure}

In the limit in which the environment is either much faster or much slower than the internal relaxation, i.e., respectively $\tau_X/\tau_D \gg 1$ and $\tau_X/\tau_D \ll 1$, we can repeat the calculations of Section I. In the presence of a slower environment, we find the following stationary joint probability distribution
\begin{align}
\label{eqn:continuous_env_joint}
    p_{\rm slow}(x_1, x_2) & = \int_{-\infty}^{+\infty} dD\, p^{\rm st}(D) p^{\rm st}(x_1, x_2|D) = \nonumber \\
    & = \frac{1}{2\pi\sqrt{(x_1^2+x_2^2)\theta \tau_D \tau_X}}e^{-\sqrt{\frac{x^2 + y^2}{\theta \tau_D \tau_X}}}
\end{align}
where $p^{\rm st}(D) \sim \mathcal{N}(0,\tau_D\theta)$ is the stationary distribution of the diffusion coefficient and $p^{\rm st}(x_1, x_2|D) \sim \mathcal{N}(0, \tau_X D^2)$ is the stationary distribution of $(x_1, x_2)$ at fixed $D$. Eq.~\eqref{eqn:continuous_env_joint} can be marginalized exactly over one of the two degrees of freedom, in order to evaluate the mutual information. The marginalization leads to
\begin{align}
\label{eqn:continuous_env_marg}
    p_{\rm slow}(x_\mu)  = \frac{1}{\pi\sqrt{\theta \tau_D \tau_X}}K_0\left(\frac{|x_\mu|}{\sqrt{\theta \tau_D \tau_X}}\right).
\end{align}
These probability distributions are plotted in Fig.~\ref{fig:continuous_env}a-b. The joint probability in Eq.~\eqref{eqn:continuous_env_joint} is not factorizable, and thus we expect the mutual information in this limit to be different from zero due to the shared environment. Since in this case there are no dimensionless parameters characterizing environmental dynamics, we expect no parametric dependence in the mutual information, $I_{\rm cont}^{\rm slow}$. Indeed, we obtain
\begin{align}
\label{eqn:continuous_env_integral}
    I_{\rm cont}^{\rm slow} = & \gamma_E + \log\frac{\pi}{2} + \\
    & - \left[1 + \frac{1}{\pi} \int_0^\infty ds\,e^{-s} \int_{0}^{2\pi} d\theta \log K_0\left(s|\cos\theta|\right)\right] \nonumber
\end{align}
where $\gamma_E$ is the Euler's constant, and the numerical value is $I_{\rm cont}^{\rm slow} \approx 0.148$.

In the opposite limit, $\tau_X/\tau_D \gg 1$, since $\ev{D}_{p^{\rm st}(D)} = 0$ we trivially find that $ p_{\rm fast}(x_1, x_2) = \prod_\mu\delta(x_\mu)$. Therefore, the mutual information vanishes in this limit. At intermediate values of $\tau_X/\tau_D$, we cannot solve Eq.~\eqref{eqn:continuous_env_FP} exactly. Therefore, to obtain samples from the stationary joint distribution, we simulate the Langevin equations Eq.~\eqref{eqn:continuous_env_langevin}. Then, from these samples, we estimate the mutual information through the k-nearest neighbors estimator proposed in \cite{kraskov2004estimating, holmes2019estimation}. The results are plot in Fig.~\ref{fig:continuous_env}c. As expected, the mutual information changes smoothly with $\tau_X/\tau_D$ and, in the limit $\tau_X/\tau_D \to 0$, approaches Eq.~\eqref{eqn:continuous_env_integral}.

\section{Conclusions}
In this work, we showed that tackling the information properties of complex systems in changing environment is a feasible task, even in the presence of non-linear interactions, non-equilibrium conditions, and continuously varying environments.

In particular, in the presence of non-linear couplings, the resulting information structure can be interpreted as an interplay between the effects of internal interactions and environmental changes. This interplay can be generically quantified by an information interference term, which surprisingly cancels exactly in the case of linear interactions \cite{nicoletti2021mutual}.

Moreover, we showed that continuously varying environments can be mapped into an effective spatial diffusion coefficient. This result might be a crucial step to understand under which conditions a shared changing environment generates effective couplings, and in which ones it does not. Additionally, the emergence of an effective space-dependent diffusion from external couplings might shed some light on the Ito/Stratonovich dilemma when describing biological and biochemical systems in inhomogeneous media.

Our results have important implications in settings where we expect non-linear or out-of-equilibrium effects to be crucial, such as neural activity originated by external stimulation \cite{neuron1, neuron2, neuron3} or population growth \cite{frey,kussell,microbial}. Notably, it was shown that, in models with latent variables, phenomenological renormalization group approaches can give seemingly non-trivial results \cite{nicoletti2020scaling, morrell2021latent}. Such models are formally similar to the framework of a changing environment analyzed here, and future works should be devoted to understand the relation between our results and the underlying information properties of these models.

Further, concepts such as mutual information and disentangled representations of the data are particularly relevant in the context of machine learning \cite{kim2018, Chen2018, Locatello2019}. Indeed, it will be paramount to unravel how these approaches might benefit from the results presented in this work.

Ultimately, we believe that this work highlights criticalities and potentialities of an information-theoretic approach to study more general and complex real-world systems. In particular, the unforeseen findings presented here might reveal, in the future, surprising properties of the information structure of complex systems with far-reaching consequences in different interdisciplinary fields.

%\bibliography{biblio}

\end{document}